\documentclass{acm_proc_article-sp}

\usepackage{graphicx}
\usepackage{epsfig, amsmath, amssymb}
\usepackage[pdftex]{hyperref}
\usepackage{algorithm}
\usepackage{algpseudocode}
\usepackage{bbm}

\newtheorem{lemma}{Lemma}

\newtheorem{theorem}[lemma]{Theorem}
\newtheorem{definition}[lemma]{Definition}

\newcommand{\bigo}{\mathcal{O}}

\begin{document}

\title{On Finding Frequent Patterns in Directed Acyclic Graphs%
\titlenote{This work was supported in part by the SPOPOS project, supported by the Danish Research and Innovation Agency under the Danish Ministry for Knowledge, Technology and Development.}}

\author{Andrea Campagna and Rasmus Pagh}

\numberofauthors{2}
\author{
\alignauthor
Andrea Campagna\\
       \affaddr{IT University of Copenhagen}\\
       \affaddr{DK-2300 K{\o}benhavn S}\\
       \affaddr{Denmark}\\
       \email{acam@itu.dk}
\alignauthor
	Rasmus Pagh\\
	       \affaddr{IT University of Copenhagen}\\
	       \affaddr{DK-2300 K{\o}benhavn S}\\
	       \affaddr{Denmark}\\
	       \email{pagh@itu.dk}
}

\maketitle

\abstract{Given a directed acyclic graph with labeled vertices, we consider the problem of finding the most common label sequences (``traces'') among all paths in the graph (of some maximum length $m$). Since the number of paths can be huge, we propose novel algorithms whose time complexity depends only on the size of the graph, and on the relative frequency $\varepsilon$ of the most frequent traces. In addition, we apply techniques from streaming algorithms to achieve space usage that depends only on $\varepsilon$, and not on the number of distinct traces.

The abstract problem considered models a variety of tasks concerning finding frequent patterns in event sequences. Our motivation comes from working with a data set of 2 million RFID readings from baggage trolleys at Copenhagen Airport. The question of finding frequent passenger movement patterns is mapped to the above problem. We report on experimental findings for this data set.
}


\section{Introduction}

Sequential pattern mining has attracted a lot of interest in recent years. However, some of the probabilistic techniques that have proven their efficiency in mining of frequent itemsets have, to our best knowledge, not been transferred to the realm of sequence mining. The aim of this paper is to take a step in that direction, namely, we propose an analogue of Toivonen's sampling-based algorithm for frequent itemset mining~\cite{toivonen} in the context of sequential patterns.

At a conceptual level we work with a new, simple formulation of the problem: The input is a directed acyclic graph (DAG) where the vertices are events and there is an edge between two events if they are considered to be connected (i.e., part of the same event sequences). Vertices are labeled by the type of event they represent. This allows certain flexibility in modeling that is lacking in many other formulations:
\begin{itemize}
\item Spatio-temporal events can be connected based on both spatial and temporal closeness.
\item Events that have an associated time range (rather than a single time stamp) can be connected based on an arbitrary closeness criterion.
\end{itemize}

The data mining task we consider is to find the most common sequences of event types (``traces'') among all paths in the DAG, or more generally all paths of some maximum length $m$. The challenge is to handle the huge number of paths that may be present in a DAG. Our approach rests on a novel sampling procedure that is able to create a sample of any desired size, in time that is linear in the size of the DAG (for preprocessing) and the size of the sample (for sampling). This allows a time complexity for the mining procedure that depends on the relative frequency $\varepsilon$ of the most common traces rather than the total number of traces. We also apply a technique from data streaming algorithms to achieve space that depends on $\varepsilon$ rather than on the number of distinct traces.

Though our formulation does not capture all the many aspects present in other approaches to sequential pattern mining, we believe that it possesses an attractive combination of {\em expressive modeling\/} and {\em algorithmic tractability}. 


\subsection{Problem definition}

We are given a directed acyclic graph $G=(V,E)$, and a function label$:V\rightarrow L$ that maps vertices to their labels. A path $p$ in $G$ is a sequence of vertices $v_1,v_2,\dots,v_j\in V$ such that $(v_i,v_{i+1})\in E$ for $i=1,\dots,j-1$.
A path $p$ has a {\em trace} $\text{label}(p)$, which is the vector of labels on the path. Let $S_m$ denote the multiset of all path traces of length at most $m$, i.e.,
$$S_m = \{ \text{label}(p) \; | \; p \text{ is a path in $G$ of length at most $m$}\} \enspace .$$
The data mining task is to find the most frequent traces in $S_m$. It comes in several flavors:
\begin{itemize}
\item {\bf Top-$k$}. For a parameter $k$, find the $k$ traces that have the most occurrences in~$S_m$ (breaking ties arbitrarily).
\item {\bf Frequency $\varepsilon$}. Find the set of traces that have relative frequency $\varepsilon$ or more in $S_m$.
\item {\bf Monte Carlo}. For both the above variants we can allow an error probability $\delta$ (typically allowing a false negative probability, i.e., that we fail to report a trace with probability $\delta$).
\end{itemize}
In this paper emphasis will be on Monte Carlo algorithms for the frequency variant. However, we note one can also obtain results for top-$k$ by a simple reduction.

\subsection{Related work}

There is a large body of related work on sequence data mining, see e.g.~\cite{journals/datamine/MannilaTV97,DBLP:conf/edbt/SrikantA96,joshi,HSDJTT,ZB_tr_2003,pisa,Chen20061203,PSpan}. These works deviate from the present one in that they consider the input as a sequence of timestamped events, and allow a host of formulations of what kinds of subsequences are of interest. 
In contrast, we put the modeling of interesting subsequences into the description of the event sequence (by defining DAG edges), and the patterns sought are simple strings. This allows us to do things that we believe have not been done, and are probably difficult, in traditional sequential data mining settings, namely making use of sampling methods. The difficulty with sampling is, of course, that patterns can overlap in complicated ways, so any straightforward approach (such as sampling nodes or edges) will fail to give independent samples.


Another related area is algorithms for finding frequent subgraphs in graphs, see e.g.~\cite{KDD'03*286,conf/kdd/HuanWPY04,journals/datamine/KuramochiK05,conf/icdm/FiedlerB07}. Indeed, the problem we consider can be seen as that of finding frequent (labeled) paths in an acyclic graph. Our work deviates from previous works mainly in that we consider directed acyclic graphs rather than general (undirected) graphs. This allows us to present algorithms with provable upper bounds on space usage and running time. No such efficient bounds are possible for general graphs: Even the problem of determining if a graph contains a simple path of length $k$ requires time exponential in $k$~\cite{JACM::AlonYZ1995,journals/ipl/Williams09}, and this is inevitable assuming the hamilton cycle problem requires exponential time in the number of vertices (a well-established hypothesis). In addition, we believe that this is the first use of sampling methods in the context of finding frequent subgraphs. Possibly, this could inspire further work on using sampling in graph mining.




\section{Our solution}

\subsection{Generation of all traces}

As a warmup we consider the task of producing the multiset of all traces having maximum length $m$.
We will use the notation $S_i(v)$ to denote the multiset of traces corresponding to  paths (of length at most $m$) starting in node $v$. Clearly $S_0(v)=\emptyset$. For $i>0$ we have the recursive definition
$$S_i(v)=\{\text{label}(v)\} \times (\epsilon \cup \bigcup_{v', (v,v')\in E} S_{i-1}(v')),$$
where $\epsilon$ denotes the empty trace, and $\bigcup$ is multiset union. Clearly we have $S_m = \bigcup_{v\in V} S_m(v)$.

These equalities lead to a simple recursive algorithm, shown in Figure~\ref{fig:naive}. It is easy to see that if traces are represented in a reasonable way (e.g.~as singly linked lists) the running time is linear in the size $|V|+|E|$ of the graph and the total length of the traces generated.

{\bf Succinct output.} If we are satisfied with returning hash values of the traces (unique with high probability) the time can be improved such that only $\bigo(1)$ time is used for each trace, i.e.~time $\bigo(|V|+|E|+|S_m|)$ in total. This can be done using a standard incremental string hashing method such as Karp-Rabin~\cite{KarpRabin}. Observe that the output is sufficient to find the {\em hash values\/} of the most frequent traces in $S_m$ (with a negligible error probability). A second run of the procedure could then output the actual frequent traces, e.g.~by looking up the count of each hash value computed.


\begin{figure}
\begin{algorithmic}[1]
\Procedure{AllTraces}{$v,t,i$}
\If{$i>0$}
\State{{\bf output} $t || \text{label}(v)$}
\For{{\bf each} $v'$ where $(v,v')\in E$}
\State{{\sc AllTraces}$(v', t || \text{label}(v),i-1)$}
\EndFor
\EndIf
\EndProcedure
\medskip
\For{$v\in V$} 
\State{{\sc AllTraces}$(v,\epsilon,m)$}
\EndFor
\end{algorithmic}
\caption{The procedure {\sc AllTraces} outputs the concatenation of a trace prefix $t$, and each trace starting at $v$ having length at most $i$. The notation $||$ is for concatenation of traces. Lines 7--9 call {\sc AllTraces} for all vertices $v$, with the empty trace $\epsilon$ as prefix, producing the multiset $S_m$ of all traces of length at most $m$.}\label{fig:naive}
\end{figure}


\subsection{Generation of a random sample}

If the patterns we are interested in occur many times, substantial savings in time can be obtained by employing a sampling procedure. That is, rather than generating $S_m$ explicitly we are interested in an algorithm that produces each trace in $S_m$ with a given probability $p$, independently. This will reduce the expected number of samples to a fraction $p$ of the original. The choice of $p$ is constrained by the fact that we still want to sample each frequent trace a fair number of times (to minimize the probability of {\em false negatives\/} being introduced by the sampling).

\paragraph{Counting phase}
Our algorithm starts by computing, for $i=1,\dots,m$ the number of paths $v.c[i]$ of length at most $i$ that start in each vertex~$v$. We assume that this can be done using standard precision (e.g.~64 bit) integers. The algorithm shown in Figure~\ref{fig:counttrace} mimics the structure of the na\"ive generation algorithm, but uses memoization (aka.~dynamic programming) to reduce the running time.

For each $i\leq m$ the cost of all calls to {\sc CountTraces} with parameters $(v,i)$, disregarding the cost of recursive calls, is easily seen to be proportional to the number of edges incident to $v$. This means that the total time complexity of the counting phase is $\bigo(|E| m)$. The space usage is dominated by an array of size $m$ for each vertex, i.e., it is $\bigo(|V| m)$.

\begin{figure}
\begin{algorithmic}[1]
\Function{CountTraces}{$v,i$}
\If{$v.c[i] = ${\bf null}}
\State{$v.c[i]\leftarrow 1$}
\For{{\bf each} $v'$ where $(v,v')\in E$}
\State{$v.c[i]\leftarrow v.c[i] +${\sc CountTraces}$(v',i-1)$}
\EndFor
\EndIf{}
\State{\Return $v.c[i]$}
\EndFunction
\medskip
\For{$v\in V$} 
\State{{\sc CountTraces}$(v,m)$}
\EndFor
\end{algorithmic}
\caption{Recursive computation of the paths of traces for each starting vertex, using memoization. It assumes that each value $v.c[0]$ is initially set to zero, and each value $v.c[i]$, $0<i\leq m$, is initially {\bf null}.}\label{fig:counttrace}
\end{figure}

\paragraph{Sampling phase}

Consider the multiset $S_i(v)$ of traces, which has size $v.c[i]$ by definition. The probability that none of these traces are sampled should be $(1-p)^{v.c[i]}$. Conditioned on the event that at least one trace from $S_i(v)$ is sampled, we either have to sample a trace of length more than one (starting with label$(v)$), or include the trace $\{v\}$ in the sample. In a nutshell, this is what the procedure {\sc SampleTraces} of Figure~\ref{fig:sampletrace} does.

Let rand() denote a function the returns a uniformly random number in $[0;1]$, independently of previously returned values. The condition $\text{rand}()>(1-p)^{v.c[m]}$ holds with probability $1-(1-p)^{v.c[m]}$, so lines 14--16 call {\sc SampleTraces}
if and only if we need to sample at least one trace from $S_m(v)$. In the procedure {\sc SampleTraces} we use, similarly to above, a parameter $t$ to pass along a trace prefix. The variable $out$ is used to keep track of whether a trace has been output in the recursive calls. If $out$ is false after all recursive calls we sample $t || \text{label}(v)$. For each $v'$ with $(v,v')\in E$ the probability that we do {\em not\/} sample any trace from $\text{label}(v) || S_{i-1}(v')$ is $(1-p)^{v'.c[i-1]}/(1-(1-p)^{v.c[i]})$. This is exactly the correct probability since we condition on at least one trace in $S_i(v)$ being sampled.

\begin{figure}
\begin{algorithmic}[1]
\Procedure{SampleTraces}{$v,t,i$}
\State{$out \leftarrow false$}
\For{{\bf each} $v'$ where $(v,v')\in E$}
\If{rand()$>(1-p)^{v'.c[i-1]}/(1-(1-p)^{v.c[i]})$}
\State{{\sc SampleTraces}$(v', t || \text{label}(v), i-1)$}
\State{$out \leftarrow true$}
\EndIf
\EndFor
\If{$out=false$ {\bf or} rand()$<p$}
\State{{\bf output} $t || \text{label}(v)$}
\EndIf
\EndProcedure
\medskip
\For{$v\in V$}
\If{rand()$>(1-p)^{v.c[m]}$}
\State{{\sc SampleTraces}$(v,\epsilon,m)$}
\EndIf
\EndFor
\end{algorithmic}
\caption{The procedure {\sc SampleTraces} outputs the concatenation of a trace prefix $t$ and a random sample of the traces starting at $v$ of length at most $i$. The traces are sampled from the conditional distribution that is guaranteed to sample at least one trace.
As before, the notation $||$ is for concatenation of traces, and $\epsilon$ denotes the empty trace. Lines 13--17 call {\sc SampleTraces} for each vertex $v$ with probability $1-(1-p)^{v.c[i]}$, to produce a sample of all traces starting at $v$ having length at most $i$, where each trace is chosen independently at random with probability~$p$.}\label{fig:sampletrace}
\end{figure}

{\bf Refinement.} Observe that the probability in line 4 may be precomputed for each edge and value of $i$. Even with this optimization, a direct implementation of the pseudocode in Figure~\ref{fig:sampletrace} may spend a lot of time in the {\bf for} loop of {\sc SampleTraces} without producing any output. To get a theoretically satisfying solution we may preprocess, for each $(v,i)$, the probabilities $p_1,p_2,\dots,p_d$ of making the recursive calls. Specifically, for $j=0,\dots,d$ we consider the probabilities $q_j=\Pi_{j'\leq j} (1-p_{j'})$ that no recursive call is made in the first $j$ iterations. If we choose $r$ uniformly at random in $[0;1]$ then the probability that $q_{j-1} > r > q_{j}$ is exactly the probability that the first recursive call is in the $j$th iteration. Similarly, the probability that $r>q_d$ is exactly the probability that no recursive call is made. Thus, by doing a binary search for $r$ over $q_d,\dots,q_0$ we may choose, with the correct probability, the first iteration $j_1$ in which there should be a recursive call. The same method can be repeated, using a random value $r$ in $[0;q_{j_1}]$ to find the next recursive call, and so on.

In the worst case this uses time $\bigo(\log |V|)$ per recursive call. We can exploit the fact that we are searching for a random value $r$ to decrease this to $\bigo(1)$ expected time. 
The basic idea is to place the probabilities $q_j$ in buckets according to the $\log d$ most significant bits, and furthermore store in each bucket its predecessor (i.e., the maximum $j$ such that $q_j$ is smaller than all elements in the bucket). 
Given $r$, we can find $j_1$ by inspecting the values in the bucket that $r$ belongs to (the elements, and their predecessor).
This will take expected time $\bigo(1)$ since $r$ is random and the average number of values per bucket is~1. 

To make this work not just for the first search, we adjust the bucketing as follows: We partition $q_1,\dots,q_d$ according to the number of leading $0$s in the binary representations (wlog.~there are $\bigo(\log n)$, since we can rely on brute-force search for low probability events, i.e., if $r$ gets very small). In each partition, containing $d'$ values, we partition the values in buckets according to the $\log d'$ most significant bits. As before, we store the predecessor of each bucket. It is clear that this data structure requires $\bigo(d)$ space, and can be constructed in time $\bigo(d)$.
A search for random $r$ in $[0;\gamma]$ happens in the structure corresponding to the number of leading $0$s in $\gamma$. This will choose a random bucket of expected size $\bigo(1)$, and the analysis finishes as before. If there are no $q_j$ values with the right number of leading $0$s, we use a special structure of $\bigo(\log n)$ bits to find the partition of the predecessor in $\bigo(1)$ time.

As before, we can choose to have a succinct output where traces are represented by the hash values of their traces, with no increase in time complexity.

\subsection{Time and error analysis}

For the time analysis we focus on the refined implementation described above, since it allows a clean and exact theoretical analysis. A similar analysis of the version stated in the pseudocode can be made under the assumption that the outdegree of vertices in $G$ is bounded by a constant. Observe that if {\sc SampleTraces} makes $c$ recursive calls this takes expected time $O(1+c)$. Also observe that the total number of procedure calls is upper bounded by the total length of all sampled traces --- this is because each recursive call is guaranteed to produce at least one output. Combining these facts we see that the expected time for all calls to {\sc SampleTraces} is linear in the length $\ell$ of all traces sampled. Notice that the expected value of $\ell$ is $\bigo(p |S_m| m)$. Since $\ell$ is independent of the random choices determining the running time of the data structure in the refined implementation we can conclude that the total expected running time of the code in Figures~\ref{fig:counttrace} and~\ref{fig:sampletrace} is $\bigo(|V| + |E| m + p |S_m| m)$.

The parameter $p$ must be chosen such that $p = C/\varepsilon$, where $C>1$ is a parameter that determines the false negative probability. The expected number of times that we sample a trace with frequency $\varepsilon'$ is $C\varepsilon'/\varepsilon$, and since the samples are independent the number of samples follows a binomial distribution. By Chernoff bounds, this means that if $\varepsilon'\geq \varepsilon$ then the number of samples is at least $C/2$ with probability $1-2^{-\Omega(C)}$. Concrete error probabilities for $C=10$ are discussed in our experimental section. We have the following theoretical result:

\begin{theorem}\label{thm:sample}
We can generate a random sample of $S_m$ in expected time $\bigo(|V|+|E|m+\log(1/\delta)/\varepsilon)$ such that each trace with frequency $\varepsilon$ or more has frequency at least $\varepsilon / 2$ in the random sample with probability $1-\delta$.
\end{theorem}

Observe that the running time is independent of the total number of traces in $S_m$.

\subsection{Putting things together}

It remains to assess how to choose, among the samples, the ones that are actually interesting. In particular, we are interested in those traces appearing in the
sample at least $C / 2$ times.

This problem can be efficiently addressed used a \textit{frequent items}
algorithm. Such algorithms have been designed for use in a data streaming
context, and guarantee low space usage.
A comprehensive treatment and an experimental comparison between various techniques can be found in~\cite{journals/pvldb/CormodeH08}.
The problem itself dates back at least to the 1980s, and can be formalized
in this way:
\begin{definition}\label{prob:freqPairs}
 Given a stream $\mathcal{S}$ of $n$ elements and a frequency threshold
 $\eta$, the frequent items problem asks for the set $\mathcal{F}$
of items that occur at least $\eta$ times.
\end{definition}
The algorithms addressing this problem usually solve a relaxed version where a modest number of false positives can appear in the output, since this reduces the space requirements to $\bigo(n/\eta)$. For completeness, we describe a concrete frequent items implementation in Appendix~\ref{app:frequent}.

In order to solve the frequent items problem without false positives,
which in our case means without reporting traces whose frequency is
below $\varepsilon$, we will make two passes, i.e., generate the sample twice and do exact counting of potentially frequent items in the second pass. This will roughly double the running time.
\begin{lemma}
 Given a stream of elements representing the set of samples of traces
 produced by \textsc{SampleTraces}, the space needed in order to output
 the traces with frequency at least $\varepsilon$, without
 producing any trace with frequency less than~$\varepsilon$, is
 $\bigo(1/\varepsilon)$ words.
\end{lemma}

Let freq$(t,S_m)$ denote $t$'s fraction of $S_m$ (viewed as a multiset). E.g., if $S_2=\{aa,aa,ab,ba,bb\}$ we have freq$(aa,S_2)=2/5$. Putting together Theorem~\ref{thm:sample} and the above lemma, we get:

\begin{theorem}
Let $\varepsilon$ and $\delta$ be positive reals.
In expected time $\bigo(|V|+|E|m+\log(1/\delta)/\varepsilon)$ and space $\bigo(1/\varepsilon)$ we can produce a set $T$ of $\bigo(1/\varepsilon)$ traces, and accompanying random variables $X_t$, $t\in T$, such that:
\begin{itemize}
\item For each $t$ with freq$(t,S_m)\geq\varepsilon$, $\Pr[t\in T] \geq 1-\delta$, and
\item for each $t\in S$, $X_t$ has binomial distribution with mean freq$(t,S_m) f(\varepsilon,\delta)$, where $f(\varepsilon,\delta)=\Theta(\log(1/\delta)/\varepsilon)$.
\end{itemize}
\end{theorem}

The first property says that the probability that a frequent trace is not reported is at most $\delta$. The second property says that the frequency of the traces in $T$ can be estimated, with strong statistical guarantees, since the $X_t$ values come from a highly concentrated distribution with mean proportional to freq$(t,S_m)$.


\section{From event sequence to a DAG}

An event sequence is a set $S$ of tuples of the form $(t,i,\ell)$, where $t\in \mathbbm{R}$ is a time stamp, $i$ is a tag identifier, and $\ell$ is a label (in our case, $\ell$ is a location identifier that indicates an approximate location, namely vicinity of an antenna). In this work we do not consider the physical locations of antenna as part of the input. 

Formally we may define the problem as follows: For a given number $\Delta$, the input set specifies a directed acyclic graph $G_\Delta = (V,E_\Delta)$, where each observation is a vertex, and there is an edge from $v_1$ to $v_2$ if and only if the vertices are observations of the same tag, at different locations, separated by at most $\Delta$ time units (we use minutes as the time unit from now on). 

To produce the DAG we sort the data by tag ID and time\-stamp. Note that this makes it easy to find all the edges from a particular vertex $v$ in $G_\Delta$: Simply scan the sorted list forward until either the timestamp differs by more than $\Delta$ from that of $v$, or we reach a node corresponding to another tag. 

{\bf Example.} If $\Delta=20$ and we observe locations 1, 2, 3, 6, 7 at time 10, 20, 30, 60, 70, the following subsequences are considered to reflect a movement: 1-2, 2-3, 1-2-3, 1-3, 6-7. Notice the inclusion of 1-3, where one observation is skipped, since there is at most $\Delta$ minutes between the observation of 1 and 3.

\subsection{Converting the RFID data}

We have worked with a data set consisting of readings of RFID (Radio-Frequency ID) tags by fixed-position antenna. RFID chips can be identified only when they are in the proximity of an antenna, which means that readings give approximate information about the location of an RFID tag. Such data sets, as well as similar data sets based on other technologies, are becoming increasingly available as more and more items, from parcels to items in shops, are being tagged with RFID chips.

Before using the RFID data to create a DAG, we have cleaned some of the noise present in the data.
One source of noise was the presence of sequences of readings
regarding trolleys remaining in zones where the range of two antennas
is overlapping.
This gave rise to sequences of alternating readings of the form $(x^+y^+)(x^+y^+)^+$.
In order to clean up these interferences, we replaced such sequences by a new zone label that represents the
zone of overlap of the range of antennas. In particular we have used,
for a sequence $(x^+y^+)(x^+y^+)^+$, the label $\min\{x,y\}*100 + \max\{x,y\}$. 
This can be thought of as an increase in the spatial resolution
of the readings.

Another source of noise, sometimes connected with the one just described,
is the presence of sequences of readings regarding the same zone for a
given trolley. In order to avoid having traces of the form $t=(Vyy^+W)$,
where $V$ and $W$ are sequences of readings, we considered only one
occurrence of $y$, properly managing the timestamps of the readings.
In particular this means that, assuming the difference
in time between any two consecutive $y$ is within the threshold $\Delta$,
in the DAG we put a directed edge $(v,y)$, $v \in V$ iff the first
occurrence of $y$ after $V$ occurred within time $\Delta$ from
$v$. Moreover we put a directed edge $(y,w)$, $w \in W$ iff $w$ happened
within time $\Delta$ from the last reading of $y$ in $t$.

\begin{figure*}
\begin{center}
\includegraphics[width=0.7\linewidth]{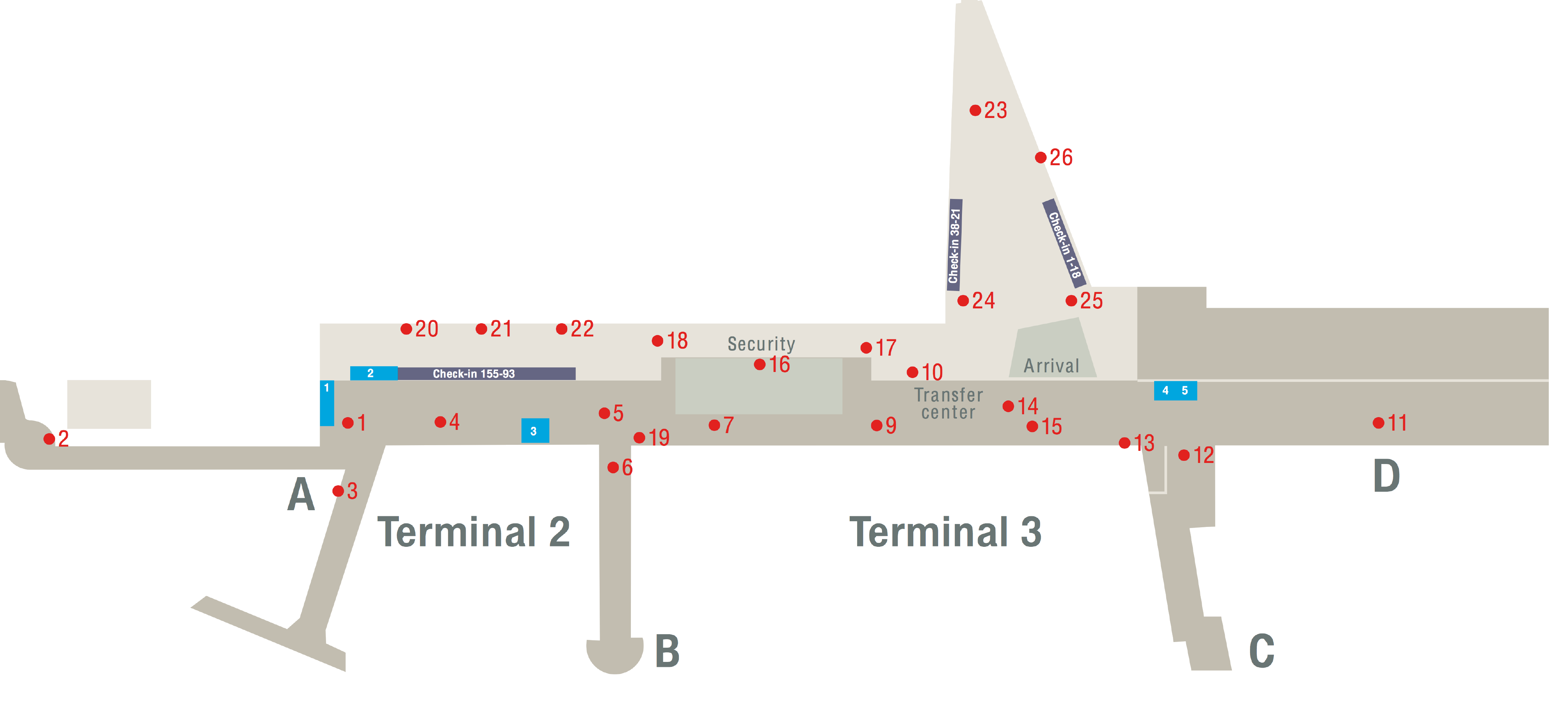}
\end{center}
\caption{RFID antenna in Copenhagen Airport. }
\end{figure*}



%
%
%


\section{Experiments}

For the experiments we have used the RFID dataset described above. We have used this dataset since it suits quite well the needs of the abstract formulation of the problem, and is massive enough to be challenging for our algorithm.
Moreover, did not manage to find interesting, raw DAG data. However, it would be of interest to try our algorithms on DAGs derived from other (publicly available) data sets.


We ran a set of experiments on the data, in order to understand how many
patterns would have been generated for a given $\Delta$ and a size $m$. Fig.~\ref{fig:results} shows the size of the graph for different sizes of $\Delta$.
\begin{figure}
 \begin{center}
  \begin{tabular}{|ccc|}
   \hline
    $\Delta$ & |V| & |E| \\
   \hline
      20 & 2206302 & 4059250\\
      10 & 2206302 & 2657931\\
       5 & 2206302 & 1721448\\
       3 & 2206302 & 1228759\\
  \hline
  \end{tabular}
\caption{Size of the airport DAG for different values of~$\Delta$. As can be seen all graphs are quite sparse, and in fact many nodes have no outgoing edges. This is due to a relatively low resolution in the data set.}
\label{fig:dagsize}
 \end{center}
\end{figure}
We compared the obtained results with the expected performance of our
algorithm (from the theoretical analysis). For space usage this gives a rather precise idea about the savings that can be obtained. For time usage, there is greater uncertainty, since the time is influenced by the constant factors in the implementation (which again depends on the hardware on which we run the experiments). It would be of interest to investigate the performance of a concrete, tuned implementation to see how close one can get to the theoretical gains.

Fig.\ref{fig:results} reports some interesting characteristics of
the data when varying $\Delta$ and $m$. In particular the table contains
the number of traces generated, the frequency of the $100$th most frequent 
trace and the ratio between the space needed in case of an
exact computation and the space required when our algorithm is used. Note that the space to represent the DAG and the counts is not taken into account in this ratio. The rationale for this is that as we consider longer event sequences the space for the DAG representation is expected to become negligible compared to the space needed for finding the most common traces. 
%
\begin{figure}
 \begin{center}
  \begin{tabular}{|cccccc|}
   \hline
    $\Delta$ & $m$ & Tot. traces & Dis. traces &top $100$th & ratio\\
   \hline
      20 & 5 & 365818472 & 4311942 & 168000 & 990\\
      10 & 5 & 106678064 & 1712646 & 52951 &  425\\
      10 & 3 & 6196850 & 50085 & 9458 & 38.2\\
       5 & 5 & 66947355 & 631300 & 42008 & 198\\
       3 & 5 & 23152990 & 280454 & 15363 & 93\\
  \hline
  \end{tabular}
\caption{Characteristics of the data for several
 combinations
 of $\Delta$ and $m$. The third column, Tot.\@ traces, represents the total
 number of traces that would be generated by the na\"\i ve approach; the
 Dis.\@ traces column represents the number of distinc traces; the top
$100$th column contains the frequency of the $100$th most frequent trace;
the column ratio represents the saving we would achive using a frequency threshold equal to the one represented in the top $100$th column.}\label{fig:results}
 \end{center}
\end{figure}

From the results of the test it is clear that great savings can be
achieved when the frequencies we are interested in are not too low.
In a case, nearly 3 orders of magnitude of space can be saved using our
approach.

Fig.~\ref{fig:time} shows the number of samples we would take in 
expectation when $C=10$ is used. The table gives the flavor of the saving
in time that could be achieved with respect to generating all the possible
traces. It is worth noticing that with $C=10$ we would end up with a probability of reporting a false positive that is lower than $7\%$ (this can be seen by considering the probability that a Poisson random variable with mean 10 or more has value less than 5). Here we notice that the total number of traces is already 1--2 orders of magnitude larger than the size of the DAG, so we expect an improvement in running time of at least 1 order of magnitude.
%
\begin{figure}
 \begin{center}
  \begin{tabular}{|ccccc|}
   \hline
    $\Delta$ & $m$ & Tot. traces & \# samples & ratio\\
   \hline
      20 & 5 & 365818472 & 22774 & 16800\\
      10 & 5 & 106678064 & 20147 & 5295\\
      10 & 3 & 6196850 & 6552 & 946\\
       5 & 5 & 66947355 & 15937 & 4200\\
       3 & 5 & 23152990 & 15070 & 1536\\
  \hline
  \end{tabular}
\caption{The ratio between the total number of traces and
 the number of samples we would take using $C=10$. Whit this value of $C$,
 the probability of having false negatives would be approximately $7\%$
 }\label{fig:time}
 \end{center}
\end{figure}



\bibliographystyle{abbrv}
\bibliography{spopos,../my}

\appendix

\section{A concrete frequent items implementation}\label{app:frequent}

For completeness, we will describe in a high level fashion
 one of the several frequent items algorithms existing in literature.
 The algorithm is presented in~\cite{KP}. We are interested in reporting
 the traces appearing at least $C/2$ times in the sample.
 For this purpose we maintain a set of $2 p |S_m| / C$ entries; each entry contains
 the label of the trace and a counter. Every time \textsc{SampleTraces}
 outputs a  trace $t$, we look at the set of entries and depending on
 whether the trace  is already recorded in one of the entries or not, we
 take one of two choices:
\begin{description}
 \item[$t$ appears in entry $i$:]we add $1$ the counter associated with the entry $i$;
 \item[$t$ does not appear in any entry:]we decrease by $1$ all the counters; if a counter reaches $0$ we remove the corresponding trace
 from the entry.
\end{description}
 This algorithm guarantees to find all the traces with frequency above
 the threshold $C/2$, but could return traces with frequency below the
 threshold. In order to eliminate this traces from the output, a second
 pass over the sample is required to get exact occurrence counts. There are two possible ways of doing this: Either one can generate exactly the same sample again (using a pseudorandom generator with the same seed, or simply by storing the random choices made). The other way (which is what we analyze theoretically) is to take a new, random sample and count exactly the number of occurrences of those elements that were found to be ``possibly frequent'' in the first sample. This increases the probability of false negatives by almost a factor of 2, so to compensate for this one needs to slightly increase $C$.

\end{document}